\documentclass[aps,twocolumn,pra, 
 superscriptaddress
]{revtex4-1} 
\usepackage{braket,amsmath,amssymb,mathrsfs,dsfont}
\usepackage{tikz}
\usetikzlibrary{matrix,decorations.pathreplacing, calc,
positioning}
\usepackage{color}
\usepackage{graphicx}
\usepackage{dcolumn}
\usepackage{bm}

\newcommand{\ie}{{\it i.~$\!$e.}}
\usepackage{xcolor}
\definecolor{maroon}{RGB}{100,20,20}
\definecolor{dblue}{RGB}{20,20,100}
\usepackage[colorlinks=true,linkcolor=dblue,citecolor=maroon,urlcolor=maroon]{hyperref}
\begin{document}
\title{Multiphoton Bell-type inequality: a tool to unearth
nonlocality of continuous variable quantum optical systems}
\author{Chandan Kumar}
\email{chandankumar@iisermohali.ac.in}
\affiliation{Department of Physical Sciences, Indian
Institute of Science Education and Research Mohali, Sector
81 SAS Nagar, Punjab 140306 India.}
\author{Gaurav Saxena}
\email{gaurav.saxena1@ucalgary.ca}
\altaffiliation[Present Address~:~]{Department of Physics and
Astronomy and Institute for Quantum Science and Technology
(IQST), University of Calgary, Calgary T2N1N4, Alberta,
Canada.}
\affiliation{Department of Physical Sciences, Indian
Institute of Science Education and Research Mohali, Sector
81 SAS Nagar, Punjab 140306 India.}
\author{Arvind}
\email{arvind@iisermohali.ac.in}
\affiliation{Department of Physical Sciences,
Indian Institute of Science Education  and
Research Mohali, Sector 81 SAS Nagar,
Punjab 140306 India.}
\begin{abstract}
We consider a multiphoton Bell-type inequality to study
nonlocality in four-mode continuous variable systems, which
goes beyond two-photon states and can be applied to mixed as
well as to states with fluctuating photon number. We apply the
inequality to a wide variety of states such as pure and
mixed Gaussian states (including squeezed thermal states)
and non-Gaussian states.  We consider beam splitters as a
model for leakage and show that the inequality is able to
detect nonlocality of noisy Gaussian states as well.
Finally, we investigate nonlocality in pair-coherent states
and entangled coherent states, which are prominent examples
of nonclassical, non-Gaussian states.
\end{abstract}
\maketitle
\section{Introduction}
\label{sec:intro}
In 1935, Albert Einstein, Boris Podolsky and Nathan Rosen in
their famous EPR paper alluded to the  possibility of the
incompleteness of Quantum Mechanics~\cite{epr}.  Bell's
seminal work of 1964 showed that attempts to complete
quantum mechanics within a local framework is
impossible~\cite{bell-1964}.  The important concepts of
entanglement and nonlocality which arose from this context
have occupied the imagination of physicists ever since and
now play a major role in the area of quantum
information~\cite{nielsen-chuang-2010,horodecki-2009}.
Violation of Bell's inequality, which is an indication of
nonlocality, is the strongest form of all quantum
correlations~\cite{nonlocality_brunner}.  In the original
EPR paper~\cite{epr}, states entangled in a continuous
degree of freedom (position) were considered.  However, most
research in nonlocality has been conducted on discrete
variable systems which involve the famous form of Bell
inequality known as the CHSH
inequality~\cite{bell2004,chsh}.  Nonlocality is useful in a
wide variety of applications such as quantum communication
and secure quantum key distribution~\cite{nl_app_1,
nl_app_2,nl_app_3,nl_app_4,nl_app_5}.  While the CHSH inequality
is sufficient for bipartite two-level
systems~\cite{nonlocality_brunner,bell2004,chsh,werner},
there have been efforts in the direction of generalizing
Bell-CHSH inequality for multipartite
systems~\cite{mpnl1,mpnl2,mpnl3,mpnl4,mpnl5,mpnl6,mpnl7}.

Formulating Bell's inequalities for Continuous Variable (CV)
systems is important as it allows us to connect with quantum
optical systems and helps us in investigating the notion of
quantumness in a variety of new situations.  Efforts have
been made to construct Bell-type inequalities for CV systems
with different number of
modes~\cite{an99,konrad-prl-1999,Chen_2002,son-prl-2006,cfrd,adesso-prl-2014}.
Specifically a generalization of the CHSH inequality for CV
systems was carried out using measurement operators having
two outcomes~\cite{an99,konrad-prl-1999,Chen_2002}. In this
formulation, modes were considered as entities, and the
analysis was not restricted to states with a fixed number of
photons.  While several studies have been performed on
pinning down nonlocality via Bell-type inequalities in
various states of the CV systems
~\cite{adesso1-pra-2017,adesso2-pra-2017,paris-pra-2004,jeong-pra-2008,nha-pra-2013},
the formulation of universal Bell-type inequalities for CV
systems  still remains an open problem.

In quantum optics, if diagonal coherent state
representation function corresponding to a quantum state is
positive and no more singular than a delta function, the
state is classified as classical, otherwise it is considered
to be nonclassical~\cite{sudarshan-prl-1963,glauber-1963}.
Classical states can be simulated by ensembles of solutions
of Maxwell equations, while nonclassical states have
intrinsic quantum properties.  The classical or nonclassical
status of a state is unaffected by the action of passive
optical elements which conserve the total photon number. On
the other hand, nonlocality  captured via Bell-type
inequalities is a consequence of quantum entanglement, which
arises in composite systems where intrinsically quantum
correlations exist.  The connection between these two
quantum features is therefore very interesting and
profound~\cite{paris-prl-2012,leuchs-pra-2015,paris-pra-2015}.
In fact there is a possibility of converting nonclassicality
into entanglement via passive
optics~\cite{paris-pra-1999,knight-pra-2002,solomon-pra-2011,
akhirpov-pra-2016,shahandeh-pra-2018}.  The notions of
classicality based on locality and optical considerations
are called C-classicality and P-classicality,
respectively~\cite{paris-pra-2015}. We demonstrate how
multiphoton  Bell-type inequalities provide an
experimentally testable connection between these two types
of nonclassicalities.

In this work, we apply the  multi-photon Bell-type
inequality~\cite{an99} to several situations in order to
demonstrate its usefulness. First, we analyze the inequality
for different two-photon states, and then consider general
pure Gaussian states. The optical circuits that we consider
convert nonclassical squeezing into
entanglement, which leads to the violation of the inequality. The
analysis of mixed states with noise is carried out for
thermal Gaussian states and for the case where dissipation
leading to loss of photons is modeled by using beam
splitters.  Nonlocality vanishes in the case of thermal states
once the temperature reaches a certain value, while
nonlocality remains preserved for all non-zero transmittance
values for the photon loss case modeled via beam splitters.
Moving beyond the class of Gaussian states, we analyze pair
coherent states and `entangled coherent states', which are
non-Gaussian nonclassical states and find that they are
indeed nonlocal and violate the multiphoton Bell-type
inequality.

This paper is organized as follows.  In
Sec.~\ref{sec:multiphotonbell}, we briefly discuss  the
multiphoton Bell violation  setup that we use in this work.
Section~\ref{subsec:twophoton} discusses
nonlocality in two-photon states,  while
Sec.~\ref{subsec:gaussian} discusses nonlocality in
four-mode general Gaussian states.
Section~\ref{subsec:nongaussian} considers non-Gaussian
states. Section~\ref{sec:conclusion} provides a summary of
our results and future directions. In
Appendix~\ref{appendix} we describe details of phase space
description of the CV systems which is used in our work.

\section{The Multiphoton Bell violation scenario}
\label{sec:multiphotonbell} 
\begin{figure}[htbp]
\includegraphics[scale=1]{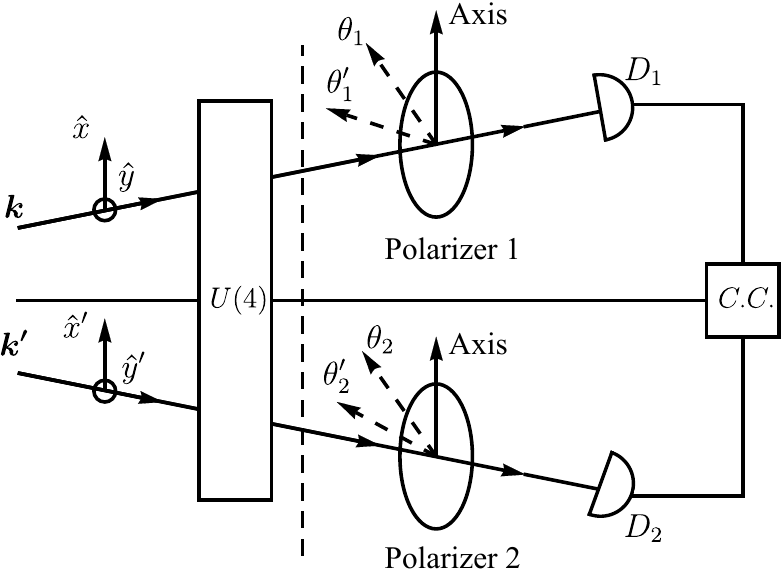}
\caption{Setup to study Bell inequality violation 
for states of a four-mode radiation field.
}
\label{fig:arvind_setup}
\end{figure}

In this section, we 
describe the setup which we consider for the violation of
Bell type inequalities.
We consider a four-mode optical system where  modes are
labeled by two wave vectors described by $ \bm{k} $ and
$\bm{k}^{\prime}$ and two polarizations are possible for
each direction as depicted in Fig.~\ref{fig:arvind_setup}.
We label the polarization basis 
by $\hat{x}$ and $\hat{y}$ 
and by
$\hat{x'}$ and $\hat{y'}$ for 
the propagation directions $\bm{k}$  and 
$\bm{k}^{\prime}$ respectively.
Quantum mechanically each mode is described by an annihilation
operator; annihilation operators $\hat{a}_1$ and $\hat{a}_2$ represent
the two polarization modes for direction $\bm{k}$, while
annihilation operators $\hat{a}_3$ and $\hat{a}_4$ correspond to the
polarization modes for the direction ${\bm k}^{\prime}$. 
We first prepare the state by applying compact passive transformations $U(4)$
consisting of beam splitters, phase shifters, and wave plates on a nonclassical
 and separable state. Subsequently, the
 photons in each propagation direction are
filtered by a polarizer placed in a particular direction to
select photons with a certain linear polarization.
After this selection, the coincidence counts are recorded
using  an on-off detector,
which performs coarse-grained measurements in the sense
it distinguishes ``light'' from ``no-light''.

We define four dichotomous Hermitian operators which
enable us to evaluate coincidence count rates and can be
used in the CHSH inequality as follows: 
\begin{eqnarray}\label{eq:operator}
\hat{A}_{1} &=& (I_{2\times2}-|00\rangle
\langle00|)_{\bm{k}}, \nonumber
\\
\hat{A}_{2} &=& (I_{2\times2}-|00\rangle
\langle00|)_{\bm{k'}}, \nonumber
\\
\hat{A}_{1}(\theta_{1}) &=& (I_{\theta_{1}} -
|0\rangle_{\theta_{1}}{}_{\theta_{1}}\langle0|)I_{\theta_{1}+\frac{\pi}{2}},
\nonumber
\\ \label{eq:arvind_operators}
\hat{A}_{2}(\theta_{2}) &=& (I_{\theta_{2}} -
|0\rangle_{\theta_{2}}{}_{\theta_{2}}\langle0|)I_{\theta_{2}+\frac{\pi}{2}}. 
\end{eqnarray}
The subscripts $\theta_{1}$ and $\theta_{2}$ are the
directions of the polarizers with subscripts $1$ and $2$
denoting propagation directions $\bm{k}$ and $\bm{k}'$,
respectively.  The quantum
mechanical action of the polarizer has been
implemented in the definition of these operators. 
The operators $\hat{A}_{1}$ and $\hat{A}_{1}(\theta_{1})$ 
act on the Hilbert space of modes $\hat{a_1}$ and $\hat{a_2}$.
The expectation value of $\hat{A}_{1}$ is the probability
of finding at least one photon with no polarizer placed 
in the path, while the expectation value of $\hat{A}_{1}(\theta_{1})$ 
is the probability of finding at least one photon after
a polarizer has been placed in the path.  
The operators $A_2$ and $A_2(\theta_2)$ play a
similar role for the modes $\hat{a_3}$ and $\hat{a_4}$.

 We now define four different types of coincidence count rates
based on different settings of the two polarizers as follows:
\begin{enumerate}
\item[(i)]
$P(\theta_{1},\theta_{2}) = \langle
\hat{A}_{1}(\theta_{1})\hat{A}_{2}(\theta_{2}) \rangle$ := The first
polarizer at $\theta_{1}$ and the second one at
$\theta_{2}$ with respect to their respective $x$
axes.
\item[(ii)]
$P(\theta_{1},\hspace{0.2cm}) =  \langle
\hat{A}_{1}(\theta_{1})\hat{A}_{2} \rangle$ := The
first polarizer at $\theta_{1}$ and the second one
removed.
\item[(iii)]
$P(\hspace{0.2cm},\theta_{2}) = \langle
\hat{A}_{1}\hat{A}_{2}(\theta_{2}) \rangle$ :=
The first polarizer  removed and the second one at
$\theta_{2}$.
\item[(iv)]
$P(\hspace{0.2cm},\hspace{0.2cm}) = \langle \hat{A}_{1}\hat{A}_{2} \rangle$ := Both the
polarizers removed from the setup.
\end{enumerate}
If the quantum state of the four mode field is known, then
the  above coincidence count rates can be readily evaluated.

If we assume that there is  local hidden variable
model (LHVM)
which can explain the outcomes of measurement of operators
given in Eq.~(\ref{eq:operator}), the coincident count
rates  have to satisfy
following inequality~\cite{chsh74} 
\begin{multline}\label{eq:multiphotoninequality}
    -P(\hspace{0.2cm},\hspace{0.2cm})\leq
P(\theta_{1},\theta_{2}) - P(\theta_{1},\theta_{2}') +
P(\theta_{1}',\theta_{2})\\
    + P(\theta_{1}',\theta_{2}') -
P(\theta_{1}',\hspace{0.2cm}) - P(\hspace{0.2cm},\theta_{2})
\leq 0.
\end{multline}
This is the state independent Bell-type inequality valid for
general radiation
states and its violation by a given state proves  that the
state has nonlocal quantum correlations that cannot be
accommodated in realist hidden variable models based on
locality. It is worth emphasizing that we have used
operators defined on the four mode field and did not imagine
photon as a single particle moving along a trajectory. 
In fact the
states that we encounter may not even have fixed number of
photons. More details regarding this inequality is available
in~\cite{an99}.

\section{Nonlocality using  Multiphoton
Bell-type Inequality}
\label{sec:examples}
In this section, we present our main results where we apply
the multiphoton Bell-type inequality to different  four
mode states of the optical field. We begin with two photon
states, and then consider a variety 
of four-mode Gaussian and non-Gaussian states.
\subsection{\label{subsec:twophoton}Two photon States}
We consider two examples of two-photon
states which are generated by applying
compact passive transformations comprising of beam
splitters, phase shifters, and wave plates. An arbitrary
passive transformation acting on our four mode system with
two spatial modes and each mode having two distinct
polarizations can be written as (see Eq.~(\ref{app:csd})
of Appendix~\ref{appendix} for more
details):
\begin{equation}\label{csd}
U = \begin{pmatrix}
U_1&0\\0&U_2
\end{pmatrix} \underbrace{\begin{pmatrix}
C&S\\-S&C
\end{pmatrix}}_D \begin{pmatrix}
V_1^T&0\\0&V_2^T
\end{pmatrix}.
\end{equation} 
To generate the  first state $|\psi_{1}\rangle$, we apply
the $U$ transformation (\ref{csd}), with
\begin{equation}
U_1 = U_2 = \frac{1}{\sqrt{2}}\begin{pmatrix}
1&-1\\1&1\\
\end{pmatrix},\,\,V_1=V_2=\mathds{1}_2, \,\, D=\mathds{1}_4,
\end{equation}
on a nonclassical and separable state:
\begin{eqnarray}
&&|0\rangle_{1}|1\rangle_{2}|0\rangle_{3}|1\rangle_{4}
\!\!
\xrightarrow{\mathcal{U}(U_1)\otimes \mathcal{U}( U_2)}
\!\!
\frac{1}{2}
(\ket{01} - \ket{10})_{12}(\ket{01}-\ket{10})_{34} 
\nonumber \\
&&\,\,\,\,\,\,\,\,\,\,\,\,\,\,\,\,=\vert \psi_1\rangle
=\frac{1}{2}(|1\rangle_{1}|0\rangle_{2}|1\rangle_{3}|0\rangle_{4}
- |1\rangle_{1}|0\rangle_{2}|0\rangle_{3}|1\rangle_{4}
\nonumber \\
&&\quad\quad\quad\quad\quad\,\,\,\,\,\,\,\,\,\,\,-
|0\rangle_{1}|1\rangle_{2}|1\rangle_{3}|0\rangle_{4} +
|0\rangle_{1}|1\rangle_{2}|0\rangle_{3}|1\rangle_{4}),
\end{eqnarray}
where $\mathcal{U}(U_1)$ and $\mathcal{U}(U_2)$ belong to  the
infinite dimensional unitary(metaplectic) representation of $U_1$ and
$U_2$ and act on the modes $1$ \& $2$ $(\bm{k})$ and modes
$3$ \& $4$ $(\bm{k^{\prime}})$, respectively.  It should be
noted that the initial state before the passive
transformation is separable and nonclassical; however, the
final state obtained after the passive transformation is
clearly entangled.  The role of passive transformations in
the generation of quantum correlations have been discussed
in Appendix~\ref{appendix}.

Similarly, the second  state  $|\psi_2 \rangle$ is  generated
 by applying the compact unitary transformation (\ref{csd}) with
 \begin{equation}
C=-S=(1/\sqrt{2})\mathds{1}_2,\,\, U_1=U_2=V_1=V_2=\mathds{1}_2,
 \end{equation}
on a nonclassical and separable state:
\begin{eqnarray}
&&    |0\rangle_{1}|0\rangle_{2}|1\rangle_{3}|1\rangle_{4}
    \xrightarrow{\mathcal{U}(D)}\frac{1}{2}(\ket{01}
    - \ket{10})_{13}(\ket{01}-\ket{10})_{24}
\nonumber \\
&&\,\,\,\,\,\,\,\,\,\,\,\,\,\,\,\,=\vert \psi_2\rangle
=\frac{1}{2}(|1\rangle_{1}|1\rangle_{2}|0\rangle_{3}|0\rangle_{4}
- |1\rangle_{1}|0\rangle_{2}|0\rangle_{3}|1\rangle_{4}
  \nonumber \\
 &&\quad\quad\quad\quad\quad\,\,\,\,\,\,\,\,\,\,\,-
|0\rangle_{1}|1\rangle_{2}|1\rangle_{3}|0\rangle_{4} +
|0\rangle_{1}|0\rangle_{2}|1\rangle_{3}|1\rangle_{4}).
\end{eqnarray}
This transformation mixes the pair of modes $1$ \& $2$
$(\bm{k})$ with the pair 
of modes $3$ \& $4$ $(\bm{k^{\prime}})$.

Explicit calculation 
shows that $\ket{\psi_{1}}$
does not violate the  multiphoton Bell-type
inequality~(\ref{eq:multiphotoninequality})
for any value of $\theta_{1}, \theta_{2},
\theta_{1}', \theta_{2}'$; however, the state
$\ket{\psi_{2}}$ does violate the inequality for
some values of $\theta_{1}, \theta_{2},
\theta_{1}', \theta_{2}'$.
In the first case since there is no entanglement between
modes belonging to two different directions, all the correlation
functions factorize, for example,
\begin{equation}\label{eq:corr}
P(\theta_{1},\theta_{2}) = \langle
\hat{A}_{1}(\theta_{1})\hat{A}_{2}(\theta_{2}) \rangle =
\langle \hat{A}_{1}(\theta_{1})\rangle \langle
\hat{A}_{2}(\theta_{2}) \rangle.
\end{equation}
Therefore, the multiphoton Bell-type inequality is obeyed.  
However, in state
$\ket{\psi_{2}}$, entanglement is present in modes $1-3$ and modes
$2-4$.  Here, unlike Eq.~(\ref{eq:corr}), 
the correlation functions,
for instance,
$\langle\hat{A}_{1}(\theta_{1})\hat{A}_{2}(\theta_{2})
\rangle \neq \langle
\hat{A}_{1}(\theta_{1})\rangle \langle \hat{A}_{2}(\theta_{2}) \rangle$,
do not factorize and this results in the violation of the
Bell-type inequality.  Thus, multiphoton Bell's
inequality~(\ref{eq:multiphotoninequality}) is designed to
detect nonlocality if entanglement exists between either of
the modes along different directions.
\subsection{Four-mode Gaussian
states}\label{subsec:gaussian}
In this section we consider various situations involving
four mode Gaussian states. We consider pure as well mixed
cases and also consider leakage modeled by beam splitters.
\subsubsection{Generic four-mode Gaussians}
\begin{figure}[htbp]
\includegraphics[scale=1]{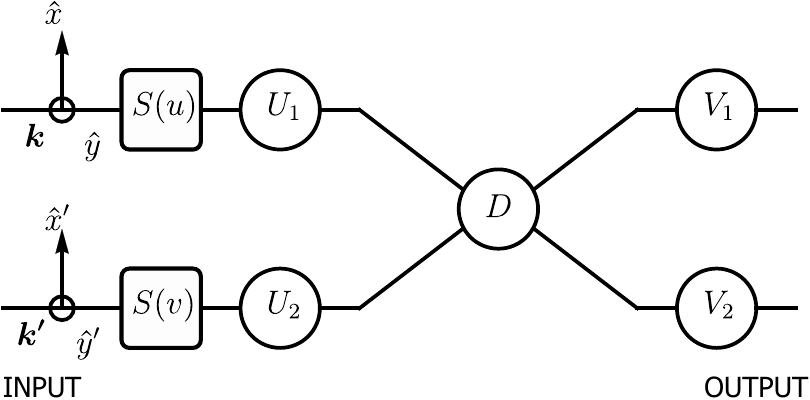}
\caption{Schematic to generate a four-mode entangled state.
Here $S(u)$ and $S(v)$ represent squeezing
transformations.  Further, $U_1$, $U_2$, $V_1$, and $V_2$
represent transformations that can be generated by
combinations of quarter and half wave plates and phase shifters, while $D$
represents transformations that can be generated using beam
splitters and quarter and half wave plates.
The first part of the circuit  generates nonclassicality by squeezing 
the individual modes and the passive operations comprising
of beam splitter, phase shifters, and quarter and half wave
 plates convert the nonclassicality into entanglement.
\newline 
$ \text{Classical}
\xrightarrow[]{\text{Squeezing}} \text{Non-classical}
\xrightarrow[]{\text{Passive operations}} \text{Entangled} $.
}
\label{fig:stategeneration}
\end{figure}
To produce a generic four-mode Gaussian state, we start with
a four-mode vacuum state or a thermal state and then apply
squeezing transformations on individual modes.  
The first
and second modes are squeezed  by an equal 
amount $u$  and  the third and fourth modes are squeezed by
an equal amount $v$. The combined symplectic
transformation corresponding to the squeezing
transformations is denoted by $S(u,v)$. The mathematical
expression for $S(u,v)$ can be readily obtained using 
Eq.~(\ref{squeezer}) given in the
Appendix~\ref{appendix} as follows:
\begin{equation}  
S(u,v)=\begin{tikzpicture}[baseline=(m-4-1.base),
every left delimiter/.style={xshift=.75em},
every right delimiter/.style={xshift=-.75em},
]
\matrix [matrix of math nodes,left delimiter=(,right delimiter=)
,row sep=0.01cm,column sep=0.01cm] (m) {
	e^{-u} & 0 & 0 & 0 & 0 & 0 & 0 & 0 \\ 
	0 & e^{-u} & 0 & 0 & 0 & 0 & 0 & 0 \\ 
	0 & 0 & e^{-v} & 0 & 0 & 0 & 0 & 0 \\ 
	0 & 0 & 0 & e^{-v} & 0 & 0 & 0 & 0 \\ 
	0 & 0 & 0 & 0 & e^{u} & 0 & 0 & 0 \\ 
	0 & 0 & 0 & 0 & 0 & e^{u} & 0 & 0 \\ 
	0 & 0 & 0 & 0 & 0 & 0 & e^{v} & 0 \\ 
	0 & 0 & 0 & 0 & 0 & 0 & 0 & e^{v}\\
};

\draw[solid] ($0.5*(m-1-4.north east)+0.5*(m-1-5.north west)$) --
($0.5*(m-8-4.south east)+0.5*(m-8-5.south west)$);
\draw[solid] ($0.5*(m-4-1.south west)+0.5*(m-5-1.north west)$) --
($0.5*(m-4-8.south east)+0.5*(m-5-8.north east)$);
\node[above=1pt of m-1-1] (top-1) {$q_1$};
\node[above=1pt of m-1-2] (top-2) {$q_2$};
\node[above=1pt of m-1-3] (top-3) {$q_3$};
\node[above=1pt of m-1-4] (top-4) {$q_4$};
\node[above=1pt of m-1-5] (top-5) {$p_1$};
\node[above=1pt of m-1-6] (top-6) {$p_2$};
\node[above=1pt of m-1-7] (top-6) {$p_3$};
\node[above=1pt of m-1-8] (top-6) {$p_4$};

\node[right=8pt of m-1-8] (left-1) {$q_1$};
\node[right=8pt of m-2-8] (left-1) {$q_2$};
\node[right=8pt of m-3-8] (left-1) {$q_3$};
\node[right=8pt of m-4-8] (left-1) {$q_4$};
\node[right=8pt of m-5-8] (left-1) {$p_1$};
\node[right=8pt of m-6-8] (left-1) {$p_2$};
\node[right=8pt of m-7-8] (left-1) {$p_3$};
\node[right=5pt of m-8-8] (left-1) {$p_4$};
\end{tikzpicture}
\end{equation}

Subsequently, the state is passed through a particular 
setting of beam splitter, phase shifters, and quarter and half wave
plates producing  an
entangled state as illustrated in Fig.~\ref{fig:stategeneration}.
We consider the passive transformation which generates the maximum
amount of entanglement when acting on a system with four
modes.
The corresponding matrix acting on the annihilation operators 
$\{ \hat{a}_1$, $\hat{a}_2$, $\hat{a}_3$, $\hat{a}_4 \}^T$ is given by
\begin{equation}
U=\frac{1}{2} \begin{pmatrix}
1 & -1 & -1 & 1  \\ 
1 & 1 & -1 & -1 \\ 
1 & -1 & 1 & -1 \\ 
1 & 1 & 1 & 1 \\ 
\end{pmatrix}.
\end{equation}
This can be decomposed  
in terms of submatrices using the form given in Eq.~(\ref{csd})
as follows :
\begin{equation}
\begin{aligned}
&U_1 =-U_2=\frac{1}{\sqrt{2}} \begin{pmatrix}
1&1\\-1&1\\
\end{pmatrix},\,\,V_1 =-V_2= \begin{pmatrix}
0&1\\-1&0\\
\end{pmatrix}, \,\,\text{and}\\
&C=S=\frac{1}{\sqrt{2}} \begin{pmatrix}
1&0\\0&1\\
\end{pmatrix}.
\end{aligned}
\end{equation}
Here  $U_1$, $U_2$, $V_1$, and $V_2$ represent
transformations that can be generated by combinations of
wave plates and phase shifters, while 
$D=\big(\begin{smallmatrix}
C&S\\-S&C
\end{smallmatrix}\big)$ 
can be generated using beam
splitters and wave plates.

The corresponding passive transformation acting on the
Hermitian quadrature operators $\hat{\xi}$ can be written as
follows using Eq.~(\ref{app:sxy}) given in 
Appendix~\ref{appendix}:
\begin{equation}  
K = \frac{1}{2}\begin{tikzpicture}[baseline=(m-4-1.base),
every left delimiter/.style={xshift=.75em},
every right delimiter/.style={xshift=-.75em},
]
\matrix [matrix of math nodes,left delimiter=(,right delimiter=),
row sep=0.01cm,column sep=0.01cm] (m) {
	1 & -1 & -1 & 1 & 0 & 0 & 0 & 0 \\ 
	1 & 1 & -1 & -1 & 0 & 0 & 0 & 0 \\ 
	1 & -1 & 1 & -1 & 0 & 0 & 0 & 0 \\ 
	1 & 1 & 1 & 1 & 0 & 0 & 0 & 0 \\ 
	0 & 0 & 0 & 0 & 1 & -1 & -1 & 1 \\ 
	0 & 0 & 0 & 0 & 1 & 1 & -1 & -1 \\ 
	0 & 0 & 0 & 0 & 1 & -1 & 1 & -1 \\ 
	0 & 0 & 0 & 0 & 1 & 1 & 1 & 1\\
};

\draw[solid] ($0.5*(m-1-4.north east)+0.5*(m-1-5.north west)$) --
($0.5*(m-8-4.south east)+0.5*(m-8-5.south west)$);
\draw[solid] ($0.5*(m-4-1.south west)+0.5*(m-5-1.north west)$) --
($0.5*(m-4-8.south east)+0.5*(m-5-8.north east)$);
\node[above=1pt of m-1-1] (top-1) {$q_1$};
\node[above=1pt of m-1-2] (top-2) {$q_2$};
\node[above=1pt of m-1-3] (top-3) {$q_3$};
\node[above=1pt of m-1-4] (top-4) {$q_4$};
\node[above=1pt of m-1-5] (top-5) {$p_1$};
\node[above=1pt of m-1-6] (top-6) {$p_2$};
\node[above=1pt of m-1-7] (top-6) {$p_3$};
\node[above=1pt of m-1-8] (top-6) {$p_4$};

\node[right=8pt of m-1-8] (left-1) {$q_1$};
\node[right=8pt of m-2-8] (left-1) {$q_2$};
\node[right=8pt of m-3-8] (left-1) {$q_3$};
\node[right=8pt of m-4-8] (left-1) {$q_4$};
\node[right=8pt of m-5-8] (left-1) {$p_1$};
\node[right=4pt of m-6-8] (left-1) {$p_2$};
\node[right=4pt of m-7-8] (left-1) {$p_3$};
\node[right=8pt of m-8-8] (left-1) {$p_4$};
\end{tikzpicture}
\end{equation}

We can write the covariance matrix of the final state generated
by the symplectic transformation $S= KS(u,v)$ acting on the thermal state as
\begin{equation}\label{eq:vfinal}
 V=KS(u,v)V_0S(u,v)^TK^T,
\end{equation}
where
\begin{equation}
V_0 = \frac{1}{2\kappa}\mathds{1}_{8\times8},\,\text{where} 
\, \kappa = \tanh\left(\frac{\hbar \omega}{2 k T}\right) \, \&  \, 0 \leq \kappa \leq 1,
\end{equation} is the 
four mode thermal state.  
Thus, $G = (1/2)V^{-1}$ can be  expressed as
\begin{equation}\label{gmatrix} 
G =KS(u,v)^{-1}G_{0}S(u,v)^{-1}K^{-1},
\end{equation}
where $G_{0} = \kappa \mathds{1}_{8\times8}$.
This $G$ matrix enables us to write the Wigner function  for
any given state using Eq.~(\ref{wignergaussian}) given in
 Appendix~\ref{appendix}.
To analyze the nonlocality of the four-mode generic Gaussian
state, we consider the average of Bell operator
\begin{equation}\label{eq:belloperator}
\begin{split}
f(\theta_{1},\theta_{2},\theta_{1}^{\prime},\theta_{2}^{\prime})
=& P(\theta_{1},\theta_{2})_{\text{qm}}^{\text{gauss}} -
P(\theta_{1},\theta_{2}^{\prime})_{\text{qm}}^{\text{gauss}}
\\
& \!\!\!\!\!\! + P(\theta_{1}^{\prime},\theta_{2})_{\text{qm}}^{\text{gauss}}
+ P(\theta_{1}^{\prime},\theta_{2}^{\prime})_{\text{qm}}^{\text{gauss}}\\
& \!\!\!\!\!\! - P(\theta_{1}^{\prime},\quad)_{\text{qm}}^{\text{gauss}} 
- P(\quad,\theta_{2})_{\text{qm}}^{\text{gauss}}. 
\end{split}
\end{equation}
We show the calculation for  one of the correlation
functions involved above:
\begin{equation}
\begin{split}
P(\theta_{1},\theta_{2})_{\text{qm}}^{\text{gauss}} 
= & 1-Tr(\rho |0\rangle_{\theta_1}
{}_{\theta_1}\langle0|)-Tr(\rho |0\rangle_{\theta_2}
{}_{\theta_2}\langle0|)\\
& + Tr(\rho |0\rangle_{\theta_1}
{}_{\theta_1}\langle0||0\rangle_{\theta_2}
{}_{\theta_2}\langle0|).
\end{split}
\end{equation}

The evaluation of the second term of the above expression
using Eq.~(\ref{innerp}) given in Appendix~\ref{appendix} in phase
space picture is shown below, while the  other terms
can be calculated in a similar way:
\begin{subequations}
\begin{align}\label{eq:generalcal}
&Tr(\rho |0\rangle_{\theta_1} {}_{\theta_1}\langle0|)  = 
2 \pi \int W(U(\theta_{1},0)\xi) W_{0}(q_1, p_1)d\xi, \\
& = 2\sqrt{\text{Det} (G)}\sqrt{ \text{Det}
	[U(\theta_1, 0)^{T}GU(\theta_1, 0)+e_{11}+e_{55}]^{-1}},
\end{align}
\end{subequations}
where $U(\theta_1,\theta_2)=R(\theta_1) \oplus R(\theta_2)\oplus
R(\theta_1)\oplus R(\theta_2)$ with
\begin{equation}
R(\theta) = \begin{pmatrix}
\cos \theta & -\sin \theta\\
\sin \theta & \cos \theta
\end{pmatrix},
\end{equation}
is the rotation in phase space caused by the polarizers
with phase space variables given in
Eq.~(\ref{eq:wig}) of Appendix~\ref{appendix}.
\subsubsection{Four-mode pure squeezed vacuum state}
Now we consider different Gaussian states
and analyze them using the framework developed above.
\begin{figure}[htbp]
\includegraphics[scale=1]{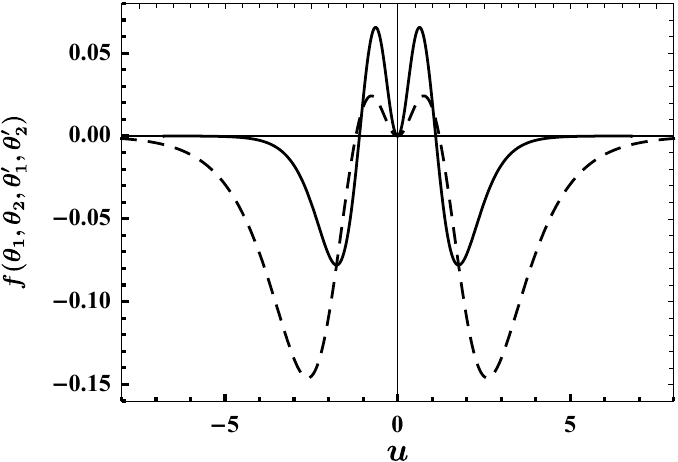}
\caption{Average of Bell operator as a function of squeezing
parameter $u$ for four-mode pure squeezed vacuum state.
Different angles are fixed as $\theta_{1} = 1.32$, 
$\theta_{2} = 0.93$, $\theta_{1}' = 3.66$, $\theta_{2}' =3.32$.
Thick solid line represents the case $v = -u$ 
corresponding to the state $\ket{\text{TMSV}}_{13}\ket{\text{TMSV}}_{24}$ 
showing violation of the  multiphoton Bell's inequality. Dashed line
corresponds to $v = 0$, which also violates the inequality
although to a less extent. 
}
\label{fig:pure4mode}
\end{figure}
We first analyze the nonlocality in four-mode pure squeezed vacuum state,
which corresponds to $\kappa =1$ in Eq.~(\ref{gmatrix}).
  Figure~\ref{fig:pure4mode} shows plot of 
  $f(\theta_{1},\theta_{2},\theta_{1}^{\prime},\theta_{2}^{\prime})$
as a function of squeezing parameter $u$
for two different cases $v = -u$,  and $v = 0$.
  Thick solid line represents the case $v = -u$ and
it violates the multiphoton Bell-type inequality. The corresponding
input state takes a very simple form
$\ket{\text{TMSV}}_{13}\ket{\text{TMSV}}_{24}$ in this case,
where TMSV denotes two mode squeezed vacuum state.
 Dashed line represents the case $v = 0$ 
and it violates the inequality
indicating that the state is nonlocal. 
 State corresponding to the case $v$ = $-u$ has the 
same entanglement structure as state $|\psi_2 \rangle$ 
which we analyzed in Sec.~\ref{subsec:twophoton}. The values
of parameters $\theta_1,\theta_2, \theta_1^{\prime}$ and
$\theta_2^{\prime}$ are chosen such that the violation of
the inequality is maximum.
\subsubsection{Four-mode squeezed thermal state}
 Thermal states of the electromagnetic field arise
when radiation is in contact with a thermal bath at a given
temperature.  We can imagine the mode under consideration
to be a classical mixture of different energy
states (states with different numbers of photons) with weight
factors given by the Boltzmann distribution.  Given a
thermal source like the Sun, if we filter out a beam along a
given direction and a fixed frequency, we will get thermal
light for the two polarisation modes.
Thermal states are classical in the quantum optical sense
and the corresponding Wigner distribution is
Gaussian. Thermal states when subjected to
squeezing transformations lead to squeezed thermal states
which again are within the class of Gaussian states however,
they are nonclassical~\cite{arvind1995}.

\begin{figure}[htbp]
\includegraphics[scale=1]{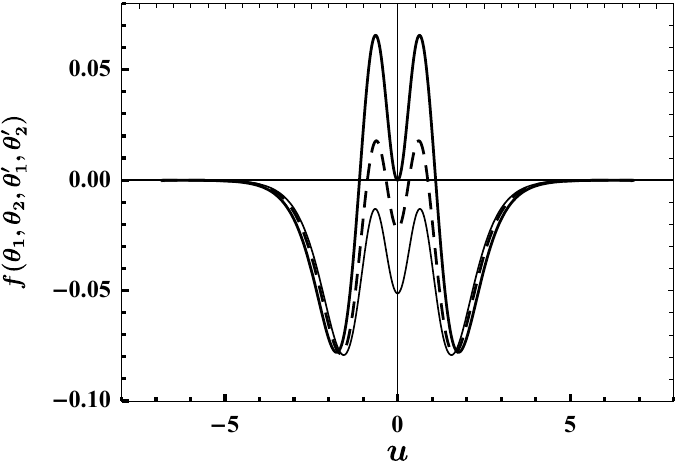} \caption{Average
of Bell operator as a function of squeezing parameter $u$
for four-mode  squeezed thermal state for the case
$v = -u$.
Different angles are fixed as $\theta_{1} = 1.32$,
$\theta_{2} = 0.93$, $\theta_{1}' = 3.66$, $\theta_{2}'
=3.32$. Thick solid, dashed, and thin solid graphs depict
$\kappa = 1$, $\kappa = 0.8$ and $\kappa = 0.7$,
respectively. Results show that an increase in temperature
results in loss of nonlocal correlations.}

\label{thermal} 
\end{figure}

We consider four-mode squeezed thermal states for the case
$v = -u$.  Figure~\ref{thermal} shows plot of
$f(\theta_{1},\theta_{2},\theta_{1}',\theta_{2}')$ as a
function of squeezing parameter $u$ for different values of
$\kappa$.  From this figure, it is clear that  nonlocal
correlations are present in the state even at a finite
temperature. As the temperature increases, the detected
nonlocal correlations vanish. It is also to be noted that
these states are nonclassical mixed states. 
\subsubsection{Leakage model}
\begin{figure}[htbp]
	\includegraphics[scale=1]{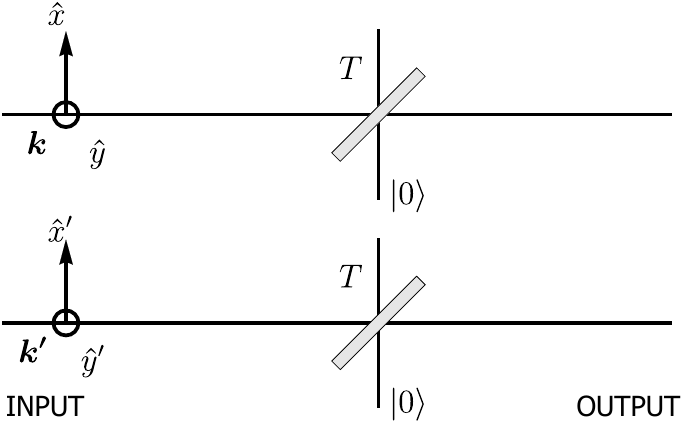}
	\caption{Modeling leakage with beam splitters.  The input
		state of the system is
		$|\text{TMSV}\rangle_{13}|\text{TMSV}\rangle_{24}$, which
		maximally violates the multi-photon Bell inequality. Each
		mode of the pure input state is mixed with vacuum using two
		beam splitters of transmittance $T$.  Subsequently, mode
		corresponding to vacuum is discarded, and thus the output is
		a mixed state.}
	\label{fig:leakage}
\end{figure}
 
We consider a scenario in which there is leakage in the
system leading to information loss and energy dissipation.
Such  leakages become quite important in various quantum
information protocols~\cite{saikat-pra-2019}, for instance,
continuous variable quantum key
distribution~\cite{radim-pra-2017}, and therefore it is
important to analyze the effects of such leakage processes
on the state properties. Typically such leakages occur due
to dissipative processes  and can be modeled with beam
splitters as shown in Fig.~\ref{fig:leakage}, where we
couple each  mode of the system in the state  $\ket{\Psi} =
\ket{\text{TMSV}}_{13}\ket{\text{TMSV}}_{24}$ with vacuum
via two beam splitters of transmittance $T$.  Subsequently,
the mode corresponding to the vacuum is traced out and the
output state of the system modes becomes a mixed Gaussian
state.


\begin{figure}[htbp]
	\begin{center}
		\includegraphics[scale=1]{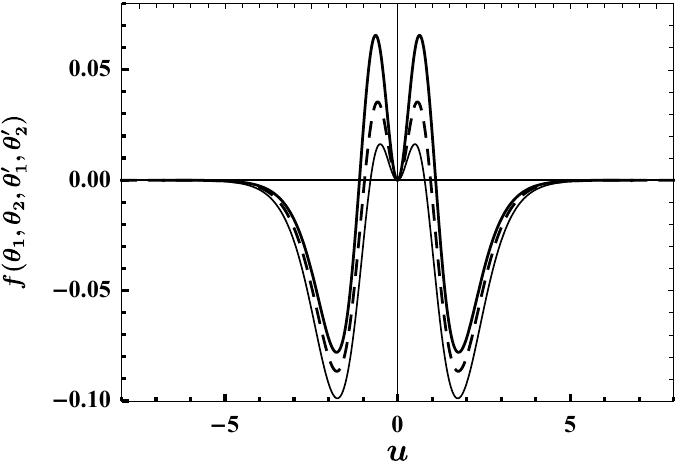}
		\caption{Average of Bell operator as a function of squeezing
			parameter $u$ for a four-mode pure squeezed vacuum state
			$\ket{\text{TMSV}}_{13}\ket{\text{TMSV}}_{24}$ in  the
			presence of leakage.  Different angles are fixed as
			$\theta_{1} = 1.32$, $\theta_{2} = 0.93$, $\theta_{1}' =
			3.66$, $\theta_{2}' =3.32$. Thick solid, dashed, and thin
			solid lines correspond to transmittance, T = 1, 0.8 and 0.6
			respectively. Results indicate that the nonlocal character
			of state remains preserved under leakage.}
		\label{fig:BSmixed}
	\end{center}
\end{figure}
The results are shown in Fig.~\ref{fig:BSmixed}.  The thick
solid, dashed and thin solid lines correspond to
transmittance T = 1, 0.8 and 0.6, respectively. We observe
that although there is a loss in the detected nonlocal
correlations as transmittance decreases, however, it never
vanishes even for low transmittance.  Hence, nonlocality of
the squeezed Gaussian state is preserved under leakage. This
is contrary to the thermal states where detected nonlocality
completely vanishes after a certain threshold temperature.


\subsection{Non-Gaussian states}\label{subsec:nongaussian}
In this section, we analyze nonlocality in families of
non-Gaussian states namely pair coherent states and
entangled coherent states. 

\subsubsection{\label{subsec:pcs1}Pair coherent states}
Pair coherent states, are a family of  non-Gaussian
entangled states of a two-mode radiation field
defined as~\cite{agarwalprl86}
\begin{equation}
\hat{a}_1 \hat{a}_2 \ket{\zeta,q} = \zeta \ket{\zeta,q}, \quad \left(\hat{a}_1
\hat{a}_1^{\dagger}-\hat{a}_2 \hat{a}_2^{\dagger}\right)\ket{\zeta,q} = q
\ket{\zeta,q}.
\label{eq:pcs}
\end{equation}
Here eigenvalue $q$ is the photon number difference between
the two-modes and eigenvalue $\zeta$ is in general complex.
Pair coherent states are simultaneous eigenkets of
$\hat{a}_1 \hat{a}_2$ and $\hat{a}_1
\hat{a}_1^{\dagger}-\hat{a}_2 \hat{a}_2^{\dagger}$.  The
solution to this eigen value problem for positive $q$ in the
Fock basis is
\begin{equation}
\ket{\zeta,q} = A_q \sum_{n=0}^{\infty}\frac{\zeta^n
}{\left[n!(n+q)!\right]^{1/2}}  \ket{n+q,n},
\end{equation}
with 
\begin{equation}
A_q = \left[\left|\zeta\right|^{-q} J_{q}\left( 2
\left|\zeta\right|\right)\right]^{-1/2}, 
\end{equation}
where $J_q$ is the modified Bessel function of the first
kind of order $q$. Entanglement, nonclassicality, and
squeezing have been studied in pair coherent
states~\cite{agarwalpcs05,agarwal88nonclassical,arvindpcs}
and these states can also be used as a resource for
teleportation~\cite{agarwaltele}.  The covariance matrix of
pair coherent states turns out to be
\begin{equation}
V(\zeta,q) = \begin{pmatrix}
N_1+\frac{1}{2}& \text{Re} \zeta&0&\text{Im} \zeta\\
\text{Re} \zeta&N_2+\frac{1}{2}& \text{Im} \zeta&0\\
0& \text{Im} \zeta&N_1+\frac{1}{2}&- \text{Re} \zeta\\
\text{Im} \zeta& 0&- \text{Re} \zeta&N_2+\frac{1}{2}
\end{pmatrix},
\end{equation}
where $N_1 = \langle \hat{a}_{1}^{\dagger}\hat{a}_{1}
\rangle$ and $N_2
= \langle \hat{a}_{2}^{\dagger}\hat{a}_{2} \rangle$.
 For non-Gaussian states, the
covariance matrix does not capture the full information,
nevertheless, studies have shown~\cite{agarwalpcs05} 
that entanglement can be detected in pair coherent state by
inequalities based on the second-order correlation. 
However, for nonlocality measurement, we cannot restrict to
Gaussian approximation of the state via the covariance
matrix.  We evaluate the
average of Bell operator~(\ref{eq:belloperator}), 
valid for general radiation states, to determine whether the
state is nonlocal or not.  Wigner function for the pair
coherent states~\cite{fan-pla-2007} can be used to calculate
the required correlation functions in phase space using
Eq.~\eqref{eq:generalcal}.

\begin{figure}[htbp]
	\begin{center}
		\includegraphics[scale=1]{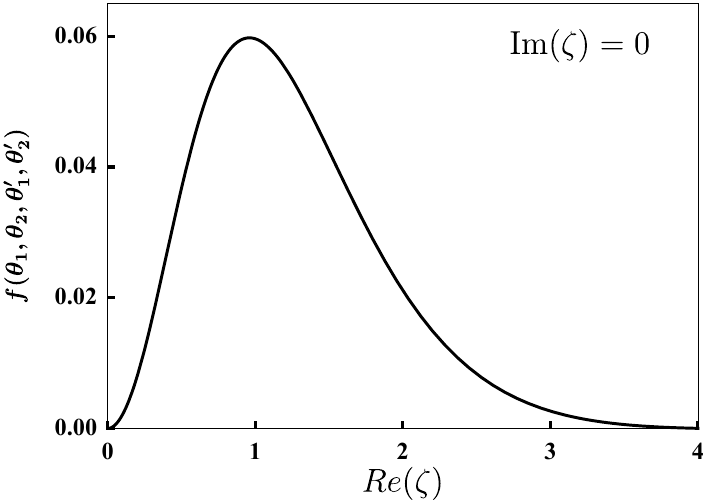}
		\caption{
			Average of Bell operator as a function of parameter
			$\text{Re}(\zeta)$ for pair coherent state with $q=0$.
			The angles are optimized to obtain maximum violation.
			The result shows that pair coherent state violates
			multiphoton Bell's inequality.}
		\label{fig:pcs}
	\end{center}
\end{figure}
We take the input state to be $\ket{PCS}_{13}\ket{PCS}_{24}$
with $q=0$ and Im($\zeta$)$=0$, and calculate the average of the
Bell operator~(\ref{eq:belloperator}). The numerically
calculated average is plotted in Fig.~\ref{fig:pcs} which
clearly shows that the family of pair coherent states
violate multiphoton Bell-type inequality.

\subsubsection{\label{subsec:ecs}Entangled Coherent State}

We consider entangled coherent state (ECS) for the two mode
system  as defined in Ref.~\cite{sanders}
\begin{equation}
|ECS\rangle = N_o\left(\left|\frac{-\alpha}{\sqrt{2}}\right\rangle
\left|\frac{\alpha}{\sqrt{2}}\right\rangle
-\left|\frac{\alpha}{\sqrt{2}}\right\rangle\left|\frac{-
\alpha}{\sqrt{2}}\right\rangle \right),
\end{equation}
where $N_{o} = [2-2\exp(-2 |\alpha|^2)]^{-1/2}$.

\begin{figure}[htbp]
	\includegraphics[scale=1]{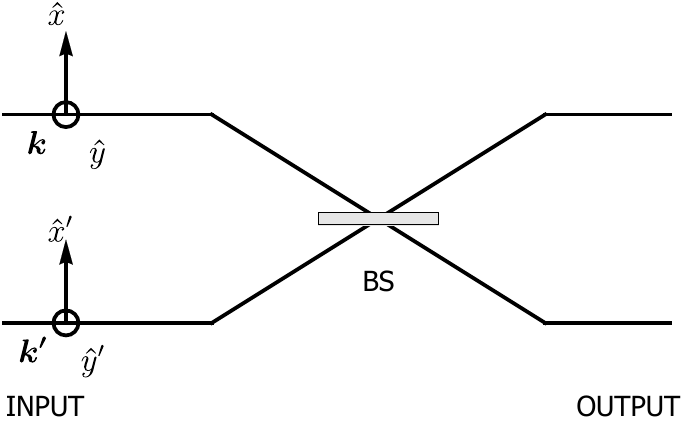}
	\caption{Set up for the generation of entangled coherent
		state (ECS). Modes $1$ and $2$ $(\bm{k})$ initialized 
		to vacuum is mixed with modes $3$ and $4$ $(\bm{k^{\prime}})$
		prepared in odd coherent state $|\psi_{o}\rangle =
		N_{o}(|\alpha \rangle-|-\alpha\rangle)$
		using a balanced beam splitter. The output state is
		$|ECS\rangle_{13}|ECS\rangle_{24}$.}
	\label{fig:ecsgen}
\end{figure}

We consider modes $1$ and $2$ (direction $\bm{k}$) initialized
to vacuum state and modes $3$ and $4$ (direction
$\bm{k^{\prime}}$)
prepared in odd coherent state	$|\psi_{o}\rangle =
N_{o}(|\alpha \rangle-|-\alpha\rangle)$. 
We then  apply the compact passive
transformation given in Eq.~(\ref{csd}) with 
\begin{equation}
C=-S=(1/\sqrt{2})\mathds{1}_2,\,\, U_1=U_2=V_1=V_2=\mathds{1}_2.
\end{equation}
This transformation  corresponds to mixing of the pair of
modes $1$ \& $2$  with the pair 
of modes $3$ \& $4$  using a balanced
beam splitter as shown in
Fig.~\ref{fig:ecsgen}, and the final state is $|ECS\rangle_{13}|ECS\rangle_{24}$: 
\begin{equation}\label{eq:ecsgeneq}
\begin{aligned}
N_{o}^2|{0}\rangle_{1}|{0}\rangle_{2}(|\alpha\rangle
-&|-\alpha\rangle)_{3}(|\alpha\rangle
-|-\alpha\rangle)_{4} 
\xrightarrow{\mathcal{U}(D)}\\
N_{o}^2 &\left(\left|\frac{-\alpha}{\sqrt{2}}\right\rangle
\left|\frac{\alpha}{\sqrt{2}}\right\rangle
-\left|\frac{\alpha}{\sqrt{2}}\right\rangle\left|\frac{-
\alpha}{\sqrt{2}}\right\rangle \right)_{13}\\
&\left(\left|\frac{-\alpha}{\sqrt{2}}\right\rangle
\left|\frac{\alpha}{\sqrt{2}}\right\rangle
-\left|\frac{\alpha}{\sqrt{2}}\right\rangle\left|\frac{-
\alpha}{\sqrt{2}}\right\rangle \right)_{24}.
\end{aligned}
\end{equation}
We use Eq.~(\ref{eq:wig}) given in Appendix~\ref{appendix} to compute the Wigner function of the
state $N_{o}^2|{0}\rangle_{1}|{0}\rangle_{2}(|\alpha\rangle
-|-\alpha\rangle)_{3}(|\alpha\rangle
-|-\alpha\rangle)_{4} $  and then transform the Wigner function as $W(\xi) \rightarrow
W\left(E^{-1} \xi\right)$ 
to obtain the Wigner function of the final state in
Eq.~\eqref{eq:ecsgeneq}, where $E$ can be written as follows
using Eq.~(\ref{app:sxy}) given in 
Appendix~\ref{appendix}:
\begin{equation}  
E = \frac{1}{\sqrt{2}}\begin{tikzpicture}[baseline=(m-4-1.base),
every left delimiter/.style={xshift=.75em},
every right delimiter/.style={xshift=-.75em},
]
\matrix [matrix of math nodes,left delimiter=(,right delimiter=),
row sep=0.01cm,column sep=0.01cm] (m) {
	1 & 0 & -1 & 0 & 0 & 0 & 0 & 0 \\ 
	0 & 1 & 0 & -1 & 0 & 0 & 0 & 0 \\ 
	1 & 0 & 1 & 0 & 0 & 0 & 0 & 0 \\ 
	0 & 1 & 0 & 1 & 0 & 0 & 0 & 0 \\ 
	0 & 0 & 0 & 0 & 1 & 0 & -1 & 0 \\ 
	0 & 0 & 0 & 0 & 0 & 1 & 0 & -1 \\ 
	0 & 0 & 0 & 0 & 1 & 0 & 1 & 0 \\ 
	0 & 0 & 0 & 0 & 0& 1 & 0 & 1\\
};

\draw[solid] ($0.5*(m-1-4.north east)+0.5*(m-1-5.north west)$) --
($0.5*(m-8-4.south east)+0.5*(m-8-5.south west)$);
\draw[solid] ($0.5*(m-4-1.south west)+0.5*(m-5-1.north west)$) --
($0.5*(m-4-8.south east)+0.5*(m-5-8.north east)$);
\node[above=1pt of m-1-1] (top-1) {$q_1$};
\node[above=1pt of m-1-2] (top-2) {$q_2$};
\node[above=1pt of m-1-3] (top-3) {$q_3$};
\node[above=1pt of m-1-4] (top-4) {$q_4$};
\node[above=1pt of m-1-5] (top-5) {$p_1$};
\node[above=1pt of m-1-6] (top-6) {$p_2$};
\node[above=1pt of m-1-7] (top-6) {$p_3$};
\node[above=1pt of m-1-8] (top-6) {$p_4$};

\node[right=8pt of m-1-8] (left-1) {$q_1$};
\node[right=8pt of m-2-8] (left-1) {$q_2$};
\node[right=8pt of m-3-8] (left-1) {$q_3$};
\node[right=8pt of m-4-8] (left-1) {$q_4$};
\node[right=8pt of m-5-8] (left-1) {$p_1$};
\node[right=4pt of m-6-8] (left-1) {$p_2$};
\node[right=8pt of m-7-8] (left-1) {$p_3$};
\node[right=8pt of m-8-8] (left-1) {$p_4$};
\end{tikzpicture}
\end{equation}

\begin{figure}[htbp]\begin{center}
\includegraphics[scale=1]{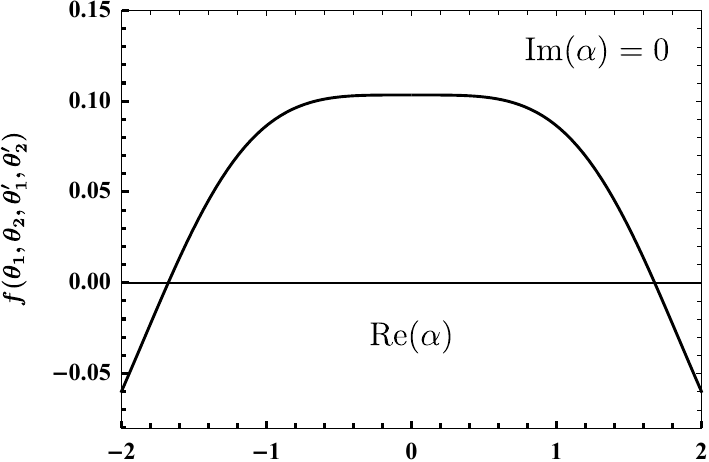}
\caption{ Average of Bell operator as a function of
$\text{Re}(\alpha)$ for entangled coherent state.
Different angles are fixed as $\theta_{1} = 2.67$,
$\theta_{2} = 5.59$, $\theta_{1}' = 1.88$, $\theta_{2}'
=3.24$.  The results clearly indicates that entangled
coherent state violates multiphoton Bell's inequality.}
\label{fig:ecsplot}
\end{center}\end{figure}

This Wigner function can be used to
compute the Bell operator in phase space, for example,
Eq.~\eqref{eq:generalcal} evaluates to
\begin{equation}
\begin{split}
Tr(\rho |0\rangle_{\theta_1} {}_{\theta_1}\langle0|)&  = 2
\frac{e^{d^2}}{(e^{d^2}-1)^2}  \bigg[-\cosh
\left(\frac{1}{4}d^2 \cos (2 \theta_1)\right)\\ &+\cosh
\left(\frac{3}{4}d^2 \right) \cosh \left(\frac{1}{4}d^2 \sin
(2 \theta_1)\right)\bigg],
\end{split}
\end{equation}
where $d= \text{Re}(\alpha)$ and $\text{Im}(\alpha)=0$.
The result is shown in Fig.~\ref{fig:ecsplot}
clearly indicating the 
violation of the multiphoton Bell-type  inequality.  

In the limit of $|\alpha|\rightarrow 0$, after expanding both sides of 
Eq.~(\ref{eq:ecsgeneq}) in the Fock basis, we obtain
\begin{equation}
|0\rangle_1|0\rangle|_2|1\rangle_{3}|1\rangle_{4}
\xrightarrow{\mathcal{U}(D)}\frac{1}{2}(|01\rangle-
|10\rangle)_{13}(|01\rangle-|10\rangle)_{24}.
\end{equation}
This is  exactly the state $|\psi_2\rangle$
considered in Sec.~(\ref{subsec:twophoton}), which has
been shown to violate the multiphoton Bell-type  inequality.
\section{Conclusion}
\label{sec:conclusion}
In this work we have explored the capacity of the
multiphoton Bell-type  inequality to unearth the
nonlocality of continuous variable systems.  In this
direction, we considered a variety of states ranging from a
finite number of photons to an arbitrary number of photons,
Gaussian to non-Gaussian.  We have used passive
transformations, which are known to convert nonclassicality
into entanglement,  to enhance the violation of the Bell-type
inequality.  The results show that the multiphoton
Bell-type  inequality, which is based on the Clauser-Horne
1974 Bell test inequality, is efficient in detecting
nonlocality in a number of situations.

The setup for the multiphoton Bell-type  inequality can
accommodate four modes.  The only requirement for the
inequality to detect nonlocality in a given state is that
the correlation should not be limited to modes 1 and 2 and
modes 3 and 4 as these pairs of modes travel along the same
physical directions.  For mixed states, the results show
that the inequality can detect nonlocality in thermal states
up to a certain temperature range.  On the other hand, when
we consider leakage modeled by beam splitters, the violation
never vanishes, 
although it diminishes with increasing leakage probability.

In our work, we have considered dichotomous measurements based
on presence of light or no light and coincidences thereof.
We are thus using very coarse grained measurements.  It
would be interesting to consider coincidence count based on
more fine grained measurements, where we distinguish
between different number of photons detected. Such
measurements are possible now and are being used and
considered in various situations~\cite{walmsley-pra-2020,chrbar-pra-2020}.
  Therefore, while  the
inequality  based on two outcomes is useful in unearthing
the nonlocality of a variety of states, finding more
general Bell-type inequalities for detecting nonlocality in
CV systems is desirable.  Another interesting direction we
are pursuing is to generalize the multiphoton Bell-type
inequality for an $n$-mode system. 
\begin{acknowledgements}
One of the authors (C. K.) thanks Sarbani Chatterjee for
encouraging discussions. A and C.K. acknowledge the financial
support from {\sf DST/ICPS/QuST/Theme-1/2019/General} Project
number {\sf Q-68}.
\end{acknowledgements}
\appendix
\section{Continuous variable system: background material}
\label{appendix} In this section, we briefly
recapitulate the CV system and their phase space
description, which have been used in our work.
\subsection{CV system and phase space}
The CV system that we consider is a four mode system as
described in Fig.~\ref{fig:arvind_setup}.  The annihilation
operators $a_j$ $(j=1,2,3, 4)$ and their conjugate creation
operators can be arranged in a column vector as
\begin{equation}
\begin{aligned}
\hat{ \xi}^{(c)} =(\xi^{(c)}_i)  = &(\hat{a_{1}}, \dots,
\hat{a_{4}}, 
\, \hat{a_{1}}^{\dagger}, \dots, \hat{a_{4}}^{\dagger})^{T},
\,\,  i = 1,2, \dots, 8.
\end{aligned}
\end{equation}
The commutation relation for the field operators can be
compactly written as
\begin{equation}\label{eq:commutcomplex}
[\hat{\xi}_i^{(c)}, \hat{\xi}_j^{(c)}] =  \beta_{ij}, \quad
\beta = \begin{pmatrix}
0_{4 } & \mathds{1}_{4 }\\
-\mathds{1}_{4 } & 0_{4}
\end{pmatrix},
\end{equation}
where $\mathds{1}_{4}$ is the $4 \times 4$ identity
matrix.  For the $i^{th}$ mode, we have the corresponding
state space spanned by the eigenvectors $\vert n_i \rangle,
\,\, \text{with} \,\, \{n_i=0,\,1, \dots ,\infty \} $ being
the corresponding eigen values of the number operator
$\hat{N}_i=\hat{a}_i^{\dagger} \hat{a}_i$.  These eigenvectors are called Fock
states or number states and the space spanned by them is the
Hilbert space $\mathcal{H}_i$ of the corresponding mode.
The combined Hilbert space $\mathcal{H}^{\otimes 4} =
\otimes_{i=1}^{4}\mathcal{H}_i$ of the four mode system is
spanned by the product basis vector $ \vert n_1\rangle\vert
n_2\rangle\vert n_3\rangle\vert n_4\rangle$ with
$\{n_1,\,n_2,\,n_3,\,n_4=0,\, 1, \dots ,\infty \} $.  The
number $n_i$ corresponds to photon number in the
$i^{\text{th}}$ mode.  The field operators $\hat{a}_i$ and
$\hat{a}^{\dagger}_i$ act irreducibly on the Hilbert space
$\mathcal{H}_i$ and their action on the number state
$|n_i\rangle$ can be easily determined by the commutation
relation given in Eq.~(\ref{eq:commutcomplex}):
\begin{equation}
\begin{aligned}
\hat{a_i}|n_i\rangle =& \sqrt{n_i}|n_i-1\rangle, \quad n_i
\geq 1,
\quad\hat{a_i}|0\rangle = 0,\\
\hat{a_i}^{\dagger}|n_i\rangle = &\sqrt{n_i+1}|n_i+1\rangle
\quad n_i \geq 0.
\end{aligned}
\end{equation}
Alternatively, we can describe our optical setup using four
pairs of Hermitian operators $\hat{q}_i,
\hat{p}_i$ for $i=1,2,3, 4$  known as quadrature operators.
These quadrature operators can be arranged in a column
vector as
\begin{equation}
\hat{ \xi} =(\hat{ \xi}_i)= (\hat{q_{1}}, \dots, \hat{q_{4}}, 
\,\hat{p_{1}}, \dots, \hat{p_{4}})^{T}, \quad i = 1,2, \dots ,8.
\end{equation}
The field operators and the quadrature operators are related as 
\begin{equation}
\hat{a}_i=   \frac{1}{\sqrt{2}}(\hat{q}_i+i\hat{p}_i), \quad 
\hat{a}^{\dagger}_i= \frac{1}{\sqrt{2}}(\hat{q}_i-i\hat{p}_i).
\end{equation}
The canonical commutation relation for the quadrature operators
can be written in a compact form as ($\hbar=1$):
\begin{equation}\label{eq:commut}
[\hat{\xi}_i, \hat{\xi}_j] =i  \beta_{ij}.
\end{equation}
The operators $\hat{q}_i$ and $\hat{p}_i$ satisfy the following 
eigenvalue equation:
\begin{equation}
\hat{q_i}|q_i\rangle = q_i|q_i\rangle, \quad 
\hat{p_i}|p_i\rangle = p_i|p_i\rangle.
\end{equation}
The eigenvalues $q_i$ and $p_i$ are real and
continuous and we have
\begin{equation}
\begin{aligned}
\langle q_i'|q_i\rangle = \delta(q_i'-q_i),& \quad 
\langle p_i'|p_i\rangle = \delta(p_i'-p_i), \\
\langle q_i|p_i\rangle =& (2 \pi)^{-1/2}e^{i q_i p_i}.
\end{aligned}
\end{equation}
\subsection{Symplectic transformations}\label{app:transformation}
The symplectic transformations for the four-mode system, which
form the non-compact group $Sp(8, \,\mathcal{R})$ are the
linear homogeneous transformations specified by real 8
$\times$ 8 matrices $S$ and they  preserve the canonical
commutation relations given in Eq.~(\ref{eq:commut})
while acting on the quadratures variables as:
\begin{equation} 
\hat{\xi}_i \rightarrow
\hat{\xi}_i^{\prime} = S_{ij}\hat{\xi}_{j}  \quad {\rm s.
t.} \quad S\beta S^T = \beta.
\end{equation}
While there are no finite dimension
unitary representations of this group, according to 
Stone-von Neumann theorem,
there exists an infinite dimensional unitary representation
$\mathcal{U}(S)$, also known as the metaplectic representation,
for each $S \in Sp(8,\, \mathcal{R})$  acting on the Hilbert
space.  For example, the metaplectic representation
$\mathcal{U}(S)$ of $S$ acts on the density operator as
$\rho \rightarrow \,\mathcal{U}(S) \rho
\,\mathcal{U}(S)^{\dagger}$.  These unitary transformations
are generated by Hamiltonians which are  quadratic functions
of quadrature and field operators.  Further, any symplectic matrix
 $S$ $\in$ $Sp(8, \,\mathcal{R})$ can be decomposed as
\begin{equation}
S = S(X,Y) P,
\end{equation}
where $S(X,Y)$ is  the maximal compact subgroup of 
$Sp(8,\, \mathcal{R})$ 
isomorphic to $U(4)$(unitary group in 4 dimensions)
and is defined as:
\begin{equation}\label{app:sxy}
S(X,Y) = 
\begin{pmatrix}
X&Y\\
-Y &X\\
\end{pmatrix},\quad X-i Y \in U(4)
\end{equation}
and $P$ $\in$ $\Pi(4)$ is a subset of  $Sp(8,\,
\mathcal{R})$ defined as
\begin{equation}
\Pi(4) = \{ S \in Sp(8, \,\mathcal{R})\,|\,S^T =S,\,\, S>0\}.
\end{equation}
In the quantum optical context, the $U(4)$ part is referred
as passive transformation and the action of its elements in
the Hilbert
space through the metaplectic representation conserve the
total photon number. Phase changes coupled with mixing via
combinations of half and quarter waves plates and beam splitters can be used to generate all
such transformations and are termed as passive operations.
Under these transformations, the classical or
nonclassical status of states does not change.  However,
such transformations have the potential to convert separable
nonclassical states into entangled nonclassical states.
On the other hand, elements of $\Pi(4)$ while acting via the
metaplectic representation do not conserve the total photon
number and are active transformations; they are also called
squeezing transformations as they can be used to generate
squeezed states. These operations can generate
nonclassicality as they can transform  a classical state to
a nonclassical one. 

The symplectic matrix for phase shift operation 
acting on the quadrature operators $\hat{q}_i$, $\hat{p}_i$ is given by
\begin{equation}
R_i(\phi) = \begin{pmatrix}
\cos \phi & \sin \phi\\
-\sin \phi & \cos \phi
\end{pmatrix}.
\end{equation}
This transformation corresponds
to $U(1)$ subgroup of $Sp(2, \mathcal{R})$.
This operation can be generated by Hamiltonian of the form 
$H =\hat{a}^{\dagger} _i\hat{a}_i$ and the corresponding 
metaplectic representation is
\begin{equation}\label{phase}
\mathcal{U}(R_i(\phi)) = 
\exp(- i \phi \underbrace{ \hat{a}^{\dagger}_i 
	\hat{a}_i}_\text{Quadratic}).  
\end{equation}

Symplectic matrix for a single mode squeezing operator
acting on quadrature operators $(\hat{q_i}, \hat{p_i})$ is
written as
\begin{equation}\label{squeezer}
S_i(u) = \begin{pmatrix}
e^{-u} & 0 \\
0 & e^{u}
\end{pmatrix}.
\end{equation}
The corresponding unitary operator acting on the Hilbert
space is given by
\begin{equation}
\mathcal{U}(S_i(u)) = 
\exp[u\underbrace{(\hat{a}_i^2-\hat{a_i}^{{\dagger}^2})}_\text{Quadratic}/2].
\label{sq_one_mode}
\end{equation}

For two-mode systems, beam splitter transformation
$B_{ij}(\theta)$
acting on quadrature operators
$  \hat{\xi} = (\hat{q}_{i}, \,\hat{q}_{j},\, \hat{p}_{i},\,
\hat{p}_{j})^{T}$ 
can be expressed as
\begin{equation}\label{beamreal}
B_{ij}(\theta) = \begin{pmatrix}
\cos\theta & -\sin\theta&0&0 \\
\sin\theta & \cos\theta&0&0\\
0 & 0&\cos\theta&-\sin\theta \\
0 & 0&\sin\theta&\cos\theta
\end{pmatrix}.
\end{equation}
The beam splitter transformation acting on field operators is
an element of the $U(2)$ compact group.
\begin{equation}\label{eq:beamc}
\begin{pmatrix}
\hat{a}_i \\
\hat{a}_j\\
\end{pmatrix}
\rightarrow
\begin{pmatrix}
\cos\theta&-\sin\theta \\
\sin\theta&\cos\theta\\
\end{pmatrix}
\begin{pmatrix}
\hat{a}_i \\
\hat{a}_j\\
\end{pmatrix}.
\end{equation}

The corresponding unitary transformation  for the  beam
splitter action is
\begin{equation}\label{metabeam}
\mathcal{U}  (B_{ij}(\theta)) =
\exp[\theta\underbrace{(\hat{a}_i \hat{a}_j^{\dagger}
-\hat{a}_i^{\dagger} \hat{a}_j)}_\text{Quadratic}].
\end{equation}
The quadratic expressions involved in Eqs.~(\ref{phase}) and (\ref{metabeam}) is
photon number conserving, while the quadratic expression
involved in Eq.~(\ref{sq_one_mode}) is not photon
conserving.  The transmittance $T$ of the beam splitter is
	 related to $\theta$ via the relation $T = \cos^2 \theta$. 
For a 50-50 (balanced) beam splitter, $\theta = \pi/4$.

Our system is comprised of two spatial modes and each spatial mode
 consists of two orthogonal polarizations, and since beam splitter
  acts only on distinct spatial modes, we also need to consider wave
   plates, which are also compact passive transformations, and can act
    on two distinct polarization modes. These wave plates along with beam
     splitters and phase shifters enable us to apply arbitrary $4 \times 4$
      compact unitary transformation on any given state. The action of 
      quarter-wave plate, whose slow axis is at an angle $\phi$ to the
       transverse direction of the electric field, on the annihilation
        operators $(\hat{a}_i,\hat{a}_j)$ is given by~\cite{mukunda-pla-1990}
\begin{equation}
Q(\phi) = \nu(\phi)C(\pi/2)\nu(\phi)^{-1},
\end{equation}
with
\begin{equation}
\nu(\phi) = \begin{pmatrix} \cos \phi&-\sin \phi\\\sin \phi& \cos \phi  \end{pmatrix}
,\,\, C(\eta) = \begin{pmatrix} e^{i\eta/2}&0\\ 0&  e^{-i\eta/2}  \end{pmatrix}.
\end{equation}
Similarly, the action of quarter-wave plate, whose slow axis is at an angle
 $\phi$ to the transverse direction of the electric field, on the annihilation
  operators $(\hat{a}_i,\hat{a}_j)$ is given by
\begin{equation}
Q(\phi) = \nu(\phi)C(\pi)\nu(\phi)^{-1}.
\end{equation}
We note that any $SU(2)$ compact transformations can be obtained as a
 combination of quarter- and half-wave plates. Further, an arbitrary
  $4 \times 4$ compact unitary transformation can be decomposed as
   following using Cosine-Sine decomposition~\cite{stewart-1982}:
\begin{equation}\label{app:csd}
U = \begin{pmatrix}
U_1&0\\0&U_2
\end{pmatrix} \underbrace{\begin{pmatrix}
C&S\\-S&C
\end{pmatrix}}_D \begin{pmatrix}
V_1^T&0\\0&V_2^T
\end{pmatrix},
\end{equation}
where  $U_1$, $U_2$, $V_1$, and $V_2$ represent $2 \times 2$ unitary transformations
that can be generated by combinations of wave plates and phase shifter, while matrix $D$, with
\begin{equation}
 C=\begin{pmatrix}
\cos \theta_1&0\\0&\cos \theta_2\end{pmatrix},\,\,S=\begin{pmatrix}
\sin \theta_1&0\\0&\sin \theta_2\end{pmatrix},
\end{equation}
 can be generated using 
beam splitters and wave plates~\cite{ish-pra-2015}.

\subsection{Phase space description}\label{app:phasespace}
The Wigner distribution 
corresponding to a density operator $\hat{\rho}$ of a
four mode quantum system is defined as
\begin{equation}\label{eq:wig}
W(\xi) = (2 \pi)^{-4}\int d^4 q'\, \langle
\underbar{q}-\frac{1}{2}
\underbar{q}^{\prime}|\hat{\rho}|\underbar{q}+\frac{1}{2}\underbar{q}^{\prime}
\rangle \exp(i \underbar{q}^{\prime}\cdot \underbar{p}),
\end{equation}
where $\underbar{q} = (q_1, q_2, q_3, q_4)^T$, $\underbar{p}
= (p_1, p_2, p_3, p_4)^T $ and $\xi = (q_{1}, \dots,
q_{4},\, p_{1}, \dots ,p_{4})^{T}$. Thus, $W(\xi)$ is a
function of eight real phase space variables  for a four mode quantum system.

First order moments  are given by
\begin{equation}
\langle \hat{\xi} \rangle = \text{Tr}[\hat{\rho}\hat{\xi}]
\end{equation} which
can be changed without affecting the quantum correlations of the state
by applying a displacement operator for the appropriate mode given by
$D(\alpha) = e^{\alpha
\hat{a_i}^{\dagger}-\alpha^{*}\hat{a_i}}$. 
The second order moments are best represented 
by the covariance matrix defined as
\begin{equation}\label{eq:cov}
V = (V_{ij})=\frac{1}{2}\langle \{\Delta \hat{\xi}_i,\Delta \hat{\xi}_j\} \rangle,
\end{equation}
where $\Delta \hat{\xi}_i = \hat{\xi}_i-\langle \hat{\xi}_i
\rangle$, and $\{\,, \, \}$ denotes an anticommutator. We note
that the covariance matrix is an 8 $\times$ 8 real, symmetric
matrix. The uncertainty principle in
terms of the covariance matrix reads
$V+\frac{i}{2}\beta \geq 0$ which
implies that the covariance matrix is positive
definite, \ie, $V>0$.

States whose Wigner distributions are Gaussian are known as
Gaussian states.  Gaussian states are completely determined
by their first and second order moments. 
We take the first order moments to be zero and thus the
covariance matrix determines the state.
The Wigner distribution
Eq.~(\ref{eq:wig}) of a general zero-centered four-mode
Gaussian state takes a simple form~\cite{arvind1995}:
\begin{equation}\label{wignergaussian}
W(q,p) =\pi^{-4}\sqrt{\text{Det} (G)}
\exp(-\xi^{T}G\xi), 
\end{equation}
where $G$ is also a real symmetric positive definite
$8\times8$ matrix related to the covariance matrix $V$ as $G =
\frac{1}{2}V^{-1}$. First order moments can always be put
back if needed, by an appropriate phase space displacement.
Coherent states, squeezed states, and thermal
states are all examples of Gaussian states and the family
contains entangled as well as non-entangled states.

Inner product of operators
$\hat{\rho}_1$ and $\hat{\rho}_2$ can be computed in 
phase space  and 
for a single mode system is given as:
\begin{equation}\label{innerp}
\text{Tr}[\hat{\rho_1}\hat{\rho_2}] = 2 \pi \int_{\mathcal{R}^2} 
dq dp\, W_{\hat{\rho}_1}(q,p) W_{\hat{\rho}_2}(q,p).
\end{equation}
\subsection{Quantum optical nonclassicality}\label{app:nonclassicality}
From a quantum optical point of view, the nonclassicality
of quantum states is defined through the Glauber-Sudarshan
representation.  Arbitrary four-mode quantum states can be
represented by the diagonal coherent state distribution
function $\phi(\bm{z})$ given by
\begin{equation}
\hat{\rho} = \frac{1}{\pi^4}\int d^8
\bm{z}\,\phi(\bm{z})|\bm{z}\rangle \langle \bm{z}|
\end{equation}
If the function $\phi(\bm{z})$ is positive and no more
singular than a delta function,  the state is defined to be
classical, otherwise  it  is  defined as nonclassical.
Coherent states and thermal states are examples of quantum
states that are classical in the above sense, whereas
quantum states such as number states, squeezed states,
superposition of coherent states are all nonclassical.

To conclude, we would like to emphasize that  all the
discussions in the above section can be generalized for an
arbitrary number of modes and details and mathematical
background is available
in~\cite{arvind-pra-1994,arvind1995,arvind-pra1995}.
%

\end{document}